\documentclass[aps,preprint]{revtex4}
\usepackage{amssymb}

\usepackage{bm}
\usepackage{amsmath}
\usepackage{graphicx}

\begin{document}

\title{$e^+e^-$ pair production in ultrarelativistic heavy-ion collisions at
intermediate impact parameters}
\author{R.N. Lee}
\email[Email:]{R.N.Lee@inp.nsk.su}
\author{A.I. Milstein}
\email[Email:]{A.I.Milstein@inp.nsk.su}
\affiliation{Budker Institute of Nuclear Physics, 630090 Novosibirsk, Russia}
\date{\today}

\begin{abstract}
Using the quasiclassical Green's function in the Coulomb field, we
analyze the probabilities of single and multiple  $e^+e^-$-pair
production at fixed impact parameter $b$ between colliding
ultrarelativistic heavy nuclei. We perform calculations in the Born
approximation with respect to the parameter $Z_B\alpha$, and exactly
in $Z_A\alpha$, $Z_{A}$ and $Z_{B}$ are the charge numbers of the
corresponding nuclei. We also obtain the approximate formulas for
the probabilities valid for $Z_A\alpha,\,Z_B\alpha\lesssim 1$.
\end{abstract}

\maketitle

\section{Introduction}

The cross section of $e^+e^-$ pair production in ultrarelativistic heavy-ion
collisions is very large, and this process can be a serious background for
many experiments. Besides, it is also important for the problem of beam
lifetime and luminosity of hadron colliders. It means that various
corrections to the Born cross section for one-pair production, as well as
the cross section for $n$-pair production ($n>1$), are very important.
Recently, the process was discussed in numerous papers , see reviews \cite%
{BB88,BHTS2002,BKN2005}. However, some important aspects of the problem have
not been finally understood, and in the present paper we are going to
elucidate them.

For our purpose, it is convenient to consider a collision of the
nuclei $A$ and $B$ with the corresponding charge numbers $Z_{A}$ and
$Z_{B}$ in the rest frame of the nucleus $A$. The nucleus $B$ is
assumed to move in the
positive direction of the $z$ axis having the Lorentz factor $\gamma$. For $%
\gamma\gg 1$, it is possible to treat the nuclei as sources of the external
field, and calculate the probability of $n$-pair production $P_n( b)$ in
collision of two nuclei at a fixed impact parameter $b$. The corresponding
cross section $\sigma_n$ is obtained by the integration over the impact
parameter,
\begin{equation}
\sigma_n=\int d^2{\ b}\,P_n( b)\, .
\end{equation}
Average number of the produced pairs at a given $b$ reads:
\begin{equation}
W( b)= \sum_{n=1}^{\infty}n P_n( b)\, .
\end{equation}
The function $W( b)$ defines the number-weighted cross section
\begin{equation}  \label{sigmaT}
\sigma_T= \int d^2{\ b}\, W( b)= \sum_{n=1}^{\infty} n\sigma_n\, .
\end{equation}
The closed expression for $\sigma_T$ was obtained in Refs.
\cite{SW,McL,Gre}, though the correct meaning of this expression was
recognized later in Ref. \cite{McL1}.

The cross section $\sigma_T$ can be presented in the form:
\begin{equation}  \label{sigmaccc}
\sigma_T=\sigma_T^0+\sigma_T^C+\sigma_T^{CC} \, ,
\end{equation}
where $\sigma_T^0$ is the Born cross section, i.e., the cross section calculated in the
lowest-order perturbation theory with respect to the parameters $Z_{A,B}\,\alpha$ (
$\sigma_T^0\propto(Z_B\alpha)^2(Z_A\alpha)^2$, $\alpha=e^2$ is the
fine-structure constant, $e$ is the electron charge, $\hbar=c=1$), $\sigma_T^C$ is the
Coulomb corrections with respect to one of the nuclei (containing the terms
proportional to $(Z_B\alpha)^{2}(Z_A\alpha)^{2n}$ or $(Z_B\alpha)^{2n}(Z_A%
\alpha)^{2}$, $n\geqslant 2$), and $\sigma_T^{CC}$ is the Coulomb
corrections with respect to both nuclei (containing the terms proportional
to $(Z_B\alpha)^{n}(Z_A\alpha)^{l}$ with $n,l> 2$). The cross section $%
\sigma_T^0$ coincides with the Born  cross section of one pair
production, which was calculated many years ago in Refs.
\cite{Landau,Racah}.

The expression for $W( b)$ derived in Refs. \cite{SW,McL,Gre}
requires regularization. The correct regularization was made in
Refs. \cite{LeeM1,LeeM2}, where the expressions for $\sigma_T^C$ and
$\sigma_T^{CC}$ were obtained in the leading logarithmic
approximation:
\begin{eqnarray}  \label{sigmaccc1}
\sigma_T^C&=&-\frac{28}{9\pi}\frac{\zeta}{m^2}\,L^2\,[f(Z_B\alpha)+f(Z_A%
\alpha)]\, ,  \notag \\
\sigma_T^{CC}&=&\frac{56}{9\pi}\frac{\zeta}{m^2}\,L\,f(Z_B\alpha)f(Z_A%
\alpha)\,,\nonumber\\
\zeta=(Z_A\alpha)^2(Z_B\alpha)^2\, ,\quad   L&=&\ln \gamma \, ,
\quad f(x)=\mathrm{Re}[\psi(1+iZ_A\alpha)+C] \,,
\end{eqnarray}
where $m$ is the electron mass, $\psi(x)=\Gamma'(x)/\Gamma(x)$, and
$C=0.577\ldots$ is the Euler constant. The expression for
$\sigma_T^C$ coincides with that obtained in Ref. \cite{Ser} by
means of the Weizs\"acker-Williams approximation. The accuracy of
the expression \eqref{sigmaccc} with $\sigma_T^C$ and
$\sigma_T^{CC}$ given in \eqref{sigmaccc1} and $\sigma_{Born}$ from
Refs. \cite{Landau,Racah}
is determined by the relative order of the omitted terms $%
\sim(Z_{A,B}\alpha)^2/L^2$. This accuracy is better than 0.4\% for
the RHIC and LHC colliders. In the recent papers
\cite{Baltz05,Baltz06}, the Coulomb corrections were calculated
numerically for a few values of $\gamma$. We emphasize that the
accuracy of the results in Refs. \cite{Baltz05,Baltz06} is the same
as in \eqref{sigmaccc1}. The uncertainty is related to the
contribution of the region, where the energies of electron and
positron are of the order of electron mass in the rest frame of one
of the nuclei.

In Refs. \cite{Baur90b,Roades91,BGS92,HTB95a} it was claimed that the
factorization of the multiple pair production probability is valid with a
good accuracy, resulting in the Poisson distribution for multiplicity:
\begin{equation}  \label{Pn}
P_n( b)=\frac{W^n( b)}{n!}\mbox{e}^{-W( b)}\,.
\end{equation}
The factor $\exp (-W)$ is nothing but the vacuum-to-vacuum
transition probability $P_0=1-\sum_{n=1}^{\infty} P_n$. Strictly
speaking, the factorization does not take place due to interference
between the diagrams corresponding to the permutation of the
electron (or positron) lines (see, e.g., \cite{McL1}). Nevertheless,
one can show that this interference gives the contribution which
contains at least one power of $L$ less than that of the amplitude
squared. Therefore, in the leading logarithmic
approximation one can use the expression \eqref{Pn}. Thus, to obtain $%
\sigma_n$ it is sufficient to know the function $W( b)$.

In Refs. \cite{HTB95b, Gucl95,Guclu00,HBT04,Guclu05}, the function
$W_0( b)$ ( the Born approximation for $W( b)$) was calculated
numerically for $m b\lesssim 1$ and a few particular values of
$\gamma$. The correct dependence of the function $W_0( b)$ on $b$ at
$m b\gg 1$ was obtained analytically in Ref. \cite{LMS02} by two
different methods. Both methods give the following result:
\begin{equation}  \label{WB1}
W_0( b) = {\frac{28}{9\pi^2}}\, {\frac{\zeta}{(m b)^2}}\; \left[2\ln{\ \gamma%
}-3\ln{( m b)}\right]\, \ln{(m b)} \, ,
\end{equation}
in the region $1\ll m b \le \sqrt{\gamma}$, and
\begin{equation}  \label{WB2}
W_0( b) = {\frac{28}{9\pi^2}}\, {\frac{\zeta}{(m b)^2}}\; \left(\ln{\frac{%
\gamma}{m b}}\right)^2\, ,
\end{equation}
in the region $\sqrt{\gamma} \le m b \ll \gamma$. Note that the function $%
W_0( b)$ given by Eqs. \eqref{WB1} and \eqref{WB2} is the continuous
function at $m b= \gamma$ together with its first derivative.
Certainly, the
integration of $W_0( b)$, Eqs. \eqref{WB1} and (%
\ref{WB2}) over $\bm{b}$ gives the main term ($\propto L^3$) in
$\sigma_T^0$. In the recent paper \cite{Guclu05}, the ansatz for
$W_0( b)$ has been suggested, that has quite different dependence of
$W_0( b)$ on $\gamma$ and $b$ for $1\ll m b \ll \sqrt{\gamma}$. In
the present paper, we confirm the result \eqref{WB1} once more and
unambiguously disprove the ansatz suggested in Ref. \cite{Guclu05}.

The cross section $\sigma_1$ of one-pair production can be represented as follows:
\begin{equation}  \label{sigma1W}
\sigma_1=\sigma_T+\sigma_{\mathrm{unit}}=\int d^2{\ b} W( b) - \int d^2{\ b}
W( b) \left(1-\mbox{e}^{-W( b)}\right) \, .
\end{equation}
Thus, the difference between $\sigma_1$ and $\sigma_T$ is due to the
unitarity correction $\sigma_{\mathrm{unit}}$. The main contribution
to the term $\sigma_T$ comes from $b\gg 1/m$. It was shown in Ref.
\cite{LMS02} that the main contribution to the second term,
$\sigma_{\mathrm{unit}}$, as well as the main contribution to the
cross sections for the n-pair production ($n\geqslant 2$), comes
from $b\sim 1/m$. As shown in Ref. \cite{LMS02},  in this region,
the function $W( b)$ has the form
\begin{equation}  \label{Eq:Ftildedef}
W( b)=\zeta\,L\,\mathcal{F}(m b)\,,
\end{equation}
where the function $\mathcal{F}(m b)$ depends on the parameters $Z_B\alpha$ and
$Z_A\alpha$ and is independent of $\gamma$. Let us represent the function
$\mathcal{F}(x)$  as
\begin{equation}\label{eq:Fexpansion}
\mathcal{F}(x)=\mathcal{F}_0(x)+\mathcal{F}_A(x)+\mathcal{F}_B(x)+\mathcal{F}_{AB}(x)\,,
\end{equation}
where $\mathcal{F}_0(x)$ is independent of $Z_A$ and $Z_B$ (Born term),
$\mathcal{F}_A(x)$ contains terms $\propto (Z_A\alpha)^{n>0}(Z_B\alpha)^{0}$ (Coulomb
corrections with respect to the nucleus $A$), $\mathcal{F}_B(x)$ contains terms $\propto
(Z_A\alpha)^0(Z_B\alpha)^{n>0}$ (Coulomb corrections with respect to the nucleus $B$),
and $\mathcal{F}_{AB}(x)$ contains terms $\propto (Z_A\alpha)^{n>0}(Z_B\alpha)^{l>0}$
(Coulomb corrections with respect to both nuclei).

In the present paper, we calculate the function $\mathcal{F}(x)$ for $%
Z_B\alpha\ll 1$ , $Z_A\alpha\lesssim 1$, and $x\lesssim 1$. In this limit, one can
neglect in Eq. \eqref{eq:Fexpansion} the terms $\mathcal{F}_B(x)$ and
$\mathcal{F}_{AB}(x)$. Though $Z_B\alpha\ll 1$, we can not expand exponent in \eqref{Pn} if $%
\zeta L\sim 1$. Our method is based on the use of the quasiclassical Green's function of
the Dirac equation in the Coulomb field.

\section{General discussion}

In the leading in $Z_B\alpha$ order, the matrix element of $e^+e^-$ pair production,
$M$, has the form
\begin{equation}
M=-e\int dt\,d\bm{r}\, \exp[-i(\varepsilon_p+\varepsilon_q) t]%
\overline{\Psi}_{p_-}(\bm{r})\,\hat{\mathcal{A}}(t,\,\bm{r}%
)\,\Psi_{-p_+}(\bm{r}),
\end{equation}
where $\mathcal{A}^\mu(t,\,\bm{r})$ is the four-vector potential of the moving nucleus
$B$, $\overline{\Psi}_{p_-}$ and $\Psi_{-p_+}$ are the positive- and negative-energy
solutions of the Dirac equation in the Coulomb field of the
nucleus $A$; $p_-=(\varepsilon_p,\,\bm{p})$, $p_+=(\varepsilon_q,\,%
\bm{q})$ are the four-momenta of electron and positron, respectively.

Then we use the Fourier transform, $\mathcal{A}^{\mu}_k$, of the vector potential $%
\mathcal{A}^\mu(t,\,\bm{r})$,
\begin{equation}
\mathcal{A}^{\mu}_k=-\frac{4\pi eZ_B}{\bm{k}_{\bot}^{2}+(k^{0}/\gamma\beta)^{2}%
}e^{-i\bm{k}_{\bot}\cdot\bm{b}}\,2\pi\delta\left( \gamma k^{0}-\gamma\beta k^{z}\right)
u^{\mu}\,,
\end{equation}
where $u^\mu=(\gamma,0,0,\gamma\beta)$ is the four-velocity of the nucleus $B$, and
$\bm{b}$ is the impact parameter. Taking the integrals over $t $, $k^0$, and $k^z$, we
obtain
\begin{align}
M & =-\frac{4\pi Z_B\alpha}{\gamma\beta}\int\frac{d\bm{k}_{\bot}}{%
\left( 2\pi\right) ^{2}}\frac{e^{-i\bm{k}_{\bot }\bm{\cdot b}%
}}{\bm{k}_{\bot}^{2}+(E/\gamma\beta)^{2}}\int d\bm{r}~\exp[i%
\bm{k}_{\bot}\bm{\cdot\rho}+iEz/\beta]\overline{\Psi}_{p_-}(%
\bm{r})\hat{u}\,\Psi_{-p_+}(\bm{r})\,,
\end{align}
where $E=\varepsilon_{p}+\varepsilon_{q}$, $\bm{r}=(\bm{\rho}%
,\,z)$.

At the calculation of the probabilities integrated over the angles of the final particles, it is
convenient to exploit the Green's functions of the
Dirac equation in the external field. Using the relations (see, e.g., \cite%
{LMS04})
\begin{align}
\sum_{\sigma }\int d\Omega _{\bm{q}}\ \Psi _{-p_{+}}(\bm{r}%
_{2})\overline{\Psi }_{-p_{+}}(\bm{r}_{1})& =-i\frac{(2\pi )^{2}}{%
q\,\varepsilon _{q}}\delta G\,(\bm{r}_{2},\bm{r}%
_{1}|-\varepsilon _{q})\,,\nonumber \\
\sum_{\sigma }\int d\Omega _{\bm{p}}\ \Psi _{p_{-}}(\bm{r}%
_{1})\overline{\Psi }_{p_{-}}(\bm{r}_{2})& =i\frac{(2\pi )^{2}}{%
p\,\varepsilon _{p}}\delta G\,(\bm{r}_{1},\bm{r}%
_{2}|\varepsilon _{p})\,,
\end{align}%
where $\delta G\,(\bm{r},\bm{r}^{\prime }|\varepsilon )$ is the discontinuity of the Green's
function on the cut, and the summation is performed over the spin states, we obtain for
the total probability:
\begin{align}
W(b)& =\sum_{\sigma _{\pm }}|M|^{2}\frac{d\bm{p}d%
\bm{q}}{(2\pi )^{6}}=\left( \frac{2Z_{B}\alpha }{\gamma \beta }%
\right) ^{2}\int \frac{d\varepsilon _{q}d\varepsilon _{p}d\bm{k}%
_{1\bot }d\bm{k}_{2\bot }}{\left( 2\pi \right) ^{4}}\frac{\exp \left[
i(\bm{k}_{1\bot }-\bm{k}_{2\bot })\bm{\cdot b}\right] }{\left[
\bm{k}_{1\bot }^{2}+(E/\gamma \beta )^{2}\right] \left[
\bm{k}_{2\bot }^{2}+(E/\gamma \beta )^{2}\right] }  \notag \\
& \times \int d\bm{r}_{1}d\bm{r}_{2}\,\exp \left[ i%
\bm{k}_{2\bot }\bm{\cdot \rho }_{2}\bm{-}i%
\bm{k}_{1\bot }\bm{\cdot \rho }_{1}+i\frac{E}{\beta }%
(z_{2}-z_{1})\right]   \notag \\
& \times \mbox{Sp}\left[ \hat{u}\,\delta G(\bm{r}_{2},\bm{r}%
_{1}|-\varepsilon _{q})\,\hat{u}\,\,\delta G(\bm{r}_{1},\bm{r%
}_{2}\,|\varepsilon _{p})\right] \,.  \label{dW1}
\end{align}%
Using the gauge invariance and the condition $\gamma \gg 1$, it is
possible to make in Eq. \eqref{dW1} the following replacement
\begin{equation}
\mbox{Sp}\left[ \hat{u}\,\delta G(\bm{r}_{2},\bm{r}%
_{1}|-\varepsilon _{q})\,\hat{u}\,\,\delta G(\bm{r}_{1},\bm{r%
}_{2}\,|\varepsilon _{p})\right] \rightarrow \frac{\gamma ^{2}}{E^{2}}\mbox{%
Sp}\left[ \hat{k}_{2\bot }\,\delta G(\bm{r}_{2},\bm{r}%
_{1}|-\varepsilon _{q})\,\hat{k}_{1\bot }\,\,\,\delta G(\bm{r}_{1},%
\bm{r}_{2}\,|\varepsilon _{p})\right] \,.
\end{equation}

In the leading logarithmic approximation, the main contribution to
the probability $W(\bm{b})$ comes from the region $\varepsilon _{\pm
}\gg m$, where the quasiclassical approximation is applicable.
Besides, it
is convenient to perform the calculations in terms of the Green's function $D(%
\bm{r},\bm{r}^{\prime }|\varepsilon )$ of the squared Dirac equation
\cite{LMS04,LMSS05}. Using the transformations similar to those in Ref. \cite{LMSS05},
we obtain
\begin{align}
W(b)& =4\left( Z_{B}\alpha \right) ^{2}\int \frac{d\varepsilon
_{q}d\varepsilon _{p}d\bm{k}_{1\bot }d\bm{k}_{2\bot }}{%
E^{2}\,\left( 2\pi \right) ^{4}}\frac{\exp \left[ i(\bm{k}_{1\bot }-%
\bm{k}_{2\bot })\bm{\cdot b}\right] }{\left[ \bm{k}%
_{1\bot }^{2}+(E/\gamma \beta )^{2}\right] \left[ \bm{k}_{2\bot
}^{2}+(E/\gamma \beta )^{2}\right] }  \notag \\
& \times \int d\bm{r}_{1}d\bm{r}_{2}\,\exp \left[ i%
\bm{k}_{2\bot }\bm{\cdot \rho }_{2}\bm{-}i%
\bm{k}_{1\bot }\bm{\cdot \rho }_{1}+i\frac{E}{\beta }%
(z_{2}-z_{1})\right]   \notag \\
& \times \mbox{Sp}\left\{ \left[ [-2i\bm{k}_{2\bot }\cdot
\bm{\nabla }_{2}+\hat{k}_{2}\hat{k}_{2\bot }]\,D(\bm{r}_{2},%
\bm{r}_{1}|-\varepsilon _{q})\right] \,\right.\nonumber\\
&\times\left. \left[ [-2i\bm{k}_{1\bot }\cdot \bm{\nabla }_{1}-\hat{k}_{1}\hat{k}_{1\bot }]\,\,D(%
\bm{r}_{1},\bm{r}_{2}\,|\varepsilon _{p})\right] \right\} \,.
\label{dW2}
\end{align}

Here $k_{1}=(E,\bm{k}_{1\bot },E)$, $k_{2}=(E,\bm{k}_{2\bot },E)$. In
the quasiclassical approximation, the function $D$ has the form
\cite{LMS04}
\begin{eqnarray}\label{Dquasi}
D(\bm{r}_{2},\bm{r}_{1}|\varepsilon )&=&\frac{i\kappa %
\mbox{e}^{i\kappa r}}{8\pi ^{2}r_{1}r_{2}}\int d\bm{q}\exp \left[ i\,%
\frac{\kappa r(\bm{q}+\bm{f})^{2}}{2r_{1}r_{2}}\right] \left( \frac{4r_{1}r_{2}}{q^2}
\right) ^{iZ_A\alpha \lambda }\nonumber\\
&&\times\left[ 1+\frac{%
\lambda r}{2r_{1}r_{2}}\bm{\alpha }\cdot \left(\bm{q}+\bm{f}\right)\right]\,,
\nonumber\\
\kappa =\sqrt{\varepsilon
^{2}-m^{2}}\,&,&\quad\lambda =\frac{\varepsilon}{\kappa} \,,\quad  \bm{\alpha }=\gamma ^{0}\bm{\gamma }%
\,,\quad \bm{f}=\frac{[[\bm{r}_{1}\times\bm{r}_{2}]\times\bm{r}]}{r^2}\,,\quad\bm{r}=\bm{r}_{1}-\bm{r}%
_{2}\,, \label{eq:Dquasiclassical}
\end{eqnarray}
where $\bm q$ is a two-dimensional vector lying in the plane perpendicular to $\bm r$.
The explicit form \eqref{Dquasi} of the quasiclassical Green's function is very convenient
for analytical investigation of high-energy processes in the Coulomb field.

\section{Analytical results}

For $mb\lesssim 1$, the main contribution to the integrals in Eq. \eqref{dW2} is given by
the region of small angles between vectors $\bm{r}_1$,  $-\bm{r}_2$, and $z$-axis. Using
these conditions and the quasiclassical Green's function \eqref{Dquasi}, we obtain the
following representation for $\mathcal{F}(mb)=\mathcal{F}_0(mb)+\mathcal{F}_A(mb) $
(details of the calculation are presented in Appendix)
\begin{equation}
\begin{split}
\mathcal{F}(mb) &  =\frac{1}{\pi^{4}\left(  Z_A\alpha\right)  ^{2}}%
\int_0^1 dx\int d^{2}Q\bm{\,}\int\frac{d^{2}\beta}{\bm{\beta }^{2}}\left[ 1-\left(
\frac{\vert\bm{R}+x\bm{Q}\vert}{\vert\bm{R}-\bar{x}\bm{Q}\vert}\right) ^{2iZ_A\alpha}\right]
\\
&  \times\left\langle 4 \sqrt{x\bar{x}}\left(  x-\bar{x}\right) %
\bm{\beta}\cdot\bm{Q}\left(
K_{1}^2(\tilde{Q})/\tilde{Q}^2-K_{1}(Q)K_{1}(\tilde{Q})/Q\tilde{Q}\right)
+\left[
K_{0}(\tilde{Q})-K_{0}(Q)\right]  ^{2}\right.  \\
&  \left.  +4x\bar{x}\beta^2
K_{1}^2(\tilde{Q})/\tilde{Q}^2+\left(  Q^{2}-4x\bar{x}\left(  \bm{\beta}%
\cdot\bm{Q}\right)  ^{2}/\beta^2\right) \,\left[
K_{1}(\tilde{Q})/\tilde{Q}-K_{1}(Q)/Q\right]  ^{2}\right\rangle ,\\
\tilde{Q}^{2} &  =Q^{2}+\beta^{2}\,,\quad\bm{R}=\sqrt{x\bar{x}%
}\bm{\beta}+m\bm{b}\,,\quad\bar{x}=1-x\,,
\end{split}\label{eq:Wres1}
\end{equation}
where $K_n(x)$ is a modified Bessel function of the third kind. The form
\eqref{eq:Wres1} is suitable for investigation of the asymptotics of $\mathcal{F}(mb)$.
For numerical evaluation, it is convenient to pass from the integration over the angle
$\phi$ of the vector $\bm{Q}$ to the integration over the parameter $v$ using the
following identities
\begin{eqnarray} &&
\int\frac{d\phi}{2\pi}\left[ 1 -\left( \frac{1+a\,\cos\phi}{1-b\,\cos\phi }\right)  ^{i\nu}\right]
\left\{
\begin{array}
[c]{c}%
1\\
\cos\phi\\
\cos2\phi
\end{array}
\right\}\nonumber\\
&&=\frac{\nu\,\sinh\pi\nu}{\pi}\lim_{\delta\to 0} \int_{0}^{1}\frac{dv}{v^{1-\delta}\,\bar{v}^{1-\delta}%
}\left(  \frac{v}{\bar{v}}\right)  ^{-i\nu}\left\{
\begin{array}
[c]{c}%
\ln\frac{1+\sqrt{1-s^{2}}}{2}\\
\frac{s}{1+\sqrt{1-s^{2}}}\\
-\frac{1}{2}\left(  \frac{s}{1+\sqrt{1-s^{2}}}\right)  ^{2}%
\end{array}
\right\} \nonumber\\
&&  \bar{v}  =1-v\,,\quad s=a\,v-b\,\bar{v}\,.%
\end{eqnarray}
Making the substitution $v=u/(u+\xi\bar{u})$, where
\begin{equation*}
\bar{u}=1-u\,,\quad
 \xi=\frac{R^{2}+\bar{x}^{2}Q^{2}}{R^{2}+x^{2}Q^{2}}\,,
\end{equation*}
and taking into account the symmetry of the integrand with respect to the substitution
$u\to \bar{u}$, $x\to \bar{x}$, we obtain

\begin{equation}
\begin{split}
\mathcal{F}(mb)  &  =4\frac{\sinh(\pi Z_A\alpha)}{\pi^4 Z_A\alpha}\int_0^1 dx\int_0^\infty dQ\,Q\bm{\,}\int\frac{d^2\beta}%
{\beta^{2}}\int_{0}^{1/2}\frac{du}{u\,\bar{u}%
}\cos\left[Z_A\alpha \ln\left( u/\bar{u}\right)  \right]\\
&  \times\left\langle \ln\left(\frac{s+\sqrt{s^2-t^2}}{g}\right)   \left\{
(1-2x\bar{x})\left[  QK_{1}(\tilde{Q})/\tilde{Q}-K_{1}(Q)\right]  ^{2}\right.\right. \\
&+\left.\left[  K_{0}%
(\tilde{Q})-K_{0}(Q)\right] ^{2}+4x\bar{x}K_{1}^2(\tilde{Q})\beta^{2}/\tilde{Q}^{2}
\right\} +x\bar{x}\left[  2\left(\frac{ \bm{\beta}\cdot\bm{R} }{\beta R}\right)^2-1\right]\\
& \times\left[\left(\frac{t}{s+\sqrt{s^{2}-t^2}}\right)^{2}-\left(\frac{R
Q\bar{x}}{g}\right)^{2}\right]\,\left[
QK_{1}(\tilde{Q})/\tilde{Q}-K_{1}(Q)\right]  ^{2} \\
& \left.%
 -4 \sqrt{x\bar{x}}\left(  \bar{x}-x\right) \frac{\bm{\beta}%
\cdot\bm{R}}{R}\,\left[\frac{t}{s+\sqrt{s^{2}-t^2}}+\frac{R Q\bar{x}}{g}\right] \left[
QK_{1}^{2}(\tilde{Q})/\tilde{Q}^{2}
-K_{1}(Q)K_{1}(\tilde{Q})\tilde{Q}\right]\right\rangle ,\\
\tilde{Q}^{2}  &  =Q^{2}+\beta^{2}\,,\quad
 g=\max(R^2,Q^2\bar{x}^2)\,, \\
t&=2Q R\left( xu-\bar{x}\bar {u}\right)\,,\quad  s =R^2+Q^2\left(
x^{2}u+\bar{x}^{2}\bar{u}\right)
 .%
\end{split}\label{eq:Wres2}
\end{equation}

Let us consider the asymptotics of Eq. \eqref{eq:Wres1}. For $mb\gg 1$, there are two
regions of integration over $\bm{\beta}$, giving the leading logarithmic contribution to
$\mathcal{F}(mb)$: $1\ll |\sqrt{x\bar{x}}\bm{\beta}+m\bm{b}|\ll mb$ and
$1\ll\beta\ll mb$. These regions give equal contributions, and the final result reads
\begin{equation}  \label{WB1a}
\mathcal{F}(mb)= \frac{56}{9\pi^2(m b)^2}\ln{(m b)} \, .
\end{equation}
Thus, the leading logarithmic contribution is given by the Born term
$\mathcal{F}_0(mb)$. This asymptotics agrees with Eq. \eqref{WB1} under the
condition $\ln (mb)\ll L$.

The main contribution to $\mathcal{F}_A(mb)$ comes from the region
$|\sqrt{x\bar{x}}\bm{\beta}+m\bm{b}|\sim 1$ and has the form
\begin{equation}\label{eq:WC1a}
\mathcal{F}_A(mb)= -\frac{28}{9\pi^2(m b)^2} f(Z_A\alpha)\, ,
\end{equation}
where the function $f(x)$ is defined in Eq. \eqref{sigmaccc1}. Again, this asymptotics is
valid under the condition $\ln (mb)\ll L$. Similar to the derivation of Eq. \eqref{WB1}, see
Ref. \cite{LMS02}, based on the equivalent photon approximation, it is possible to obtain
for the Coulomb corrections to $W(b)$ with respect to the nucleus $A$,  $W_A(b)$, the
expression valid in the wider region $\ln (mb)\lesssim L$ (but still $1\ll mb\ll \gamma$).
We have
\begin{equation}\label{eq:WC1}
W_A(b)= -\frac{28}{9\pi^2}\frac{\zeta }{(m b)^2} f(Z_A\alpha)
\ln\left(\frac{\gamma}{mb}\right)\, .
\end{equation}
 Eq. \eqref{eq:WC1a} evidently agrees with  Eq. \eqref{eq:WC1}.

 Let us consider the asymptotics at small impact parameters. For $mb\ll
 1$, the leading logarithmic contribution comes from the region $mb\ll \beta\sim Q\ll
 1$. Taking the integrals over this region, we obtain
 \begin{equation}
 \begin{split}
\mathcal{F}(mb)=&\frac{8}{3\pi^2(Z_A\alpha)^{2}} \ln\left (\frac{1}{mb}\right)
\textrm{Re}\left[
 \psi(1+i Z_A\alpha)+C-(Z_A\alpha)^2\right.\\
& \left.+iZ_A\alpha(1+(Z_A\alpha)^2)\psi'(1+iZ_A\alpha) \right]\,.
\end{split}\label{eq:smallb}
 \end{equation}
This asymptotics is obtained for zero nuclear radius $R_n$. In order to obtain $W(b)$ for
the extended nuclei, it is sufficient, within the logarithmic accuracy, to make the
substitution $\ln\left (mb\right) \to \ln\left (mb+mR_n\right)$ in the asymptotics
\eqref{eq:smallb}. For $b\gg R_n$, the finite-nuclear-size correction to $W(b)$ is
negligible.

\section{Numerical results}

\begin{figure}
 \centering\setlength{\unitlength}{0.12cm}
\begin{picture}(120,80)
 \put(32,0){\makebox(0,0)[t]{$x$}}
 \put(92,0){\makebox(0,0)[t]{$x$}}
 \put(-2,30){\rotatebox[origin=c]{90}{\makebox(0,0)[t]{$\mathcal{F}(x)$}}}
\put(0,0){\includegraphics[width=120\unitlength,keepaspectratio=true]{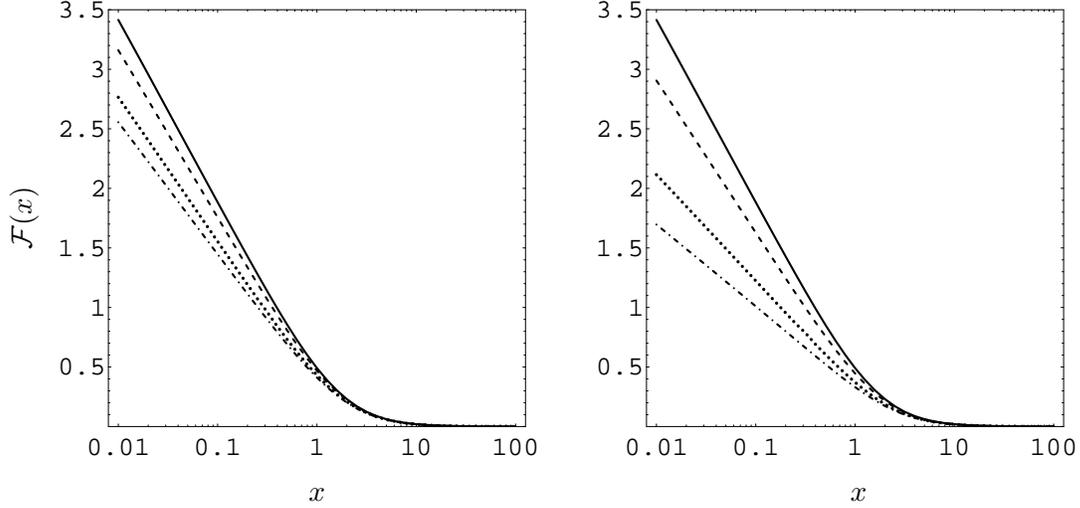}}
\end{picture}
\caption{The function $\mathcal{F}(x)$,  Eq. \eqref{Eq:Ftildedef}, for $Z_A=92$
(dash-dotted line) , $Z_A=79$ (dotted line), $Z_A=47$ (dashed line), and the Born
approximation (solid line). Left plot corresponds to the Born approximation in
$Z_B\alpha$, Eq. \eqref{eq:Wres2}. Right plot shows the results obtained from Eqs.
\eqref{eq:Fexpansion} and \eqref{eq:Wres2} for $Z_B=Z_A$ with the term
$\mathcal{F}_{AB}(x)$ omitted.}\label{fig:Ffunctions}
\end{figure}

Using Eq. \eqref{dW2}, we performed the tabulation of the function $\mathcal{F}(mb)$ for
a few values of $Z_A$. The corresponding results are presented on the left plot of Fig.
\ref{fig:Ffunctions} and in the Table \ref{tab:Ffunctions}. We remind that  these results
are obtained in the Born approximation with respect to the nucleus $B$. For most
experiments  $Z_A=Z_B$, and it is necessary to know the function $\mathcal{F}(mb)$
beyond the Born approximation with respect to the nucleus $B$. If we assume that the
term $\mathcal{F}_{AB}$ in Eq. \eqref{eq:Fexpansion} is numerically small, then we can
approximate the function $\mathcal{F}$ as $\mathcal{F}_0+2\mathcal{F}_A$ in this case.
This function is shown on the right plot of Fig. \ref{fig:Ffunctions}. It is seen that the
Coulomb corrections in the region $mb\lesssim 1$ are very important for the
experimentally interesting case $Z_A=Z_B=79$. The  assumption of smallness of the
contribution $\mathcal{F}_{AB}$ is supported by the comparison of our results for
$W(b)$ with those obtained in Refs. \cite{HTB99,Baltz06} for $Z_A=Z_B=79$ and
$\gamma=2\times10^4$ ($\gamma_{c.m.}=100$).

\begin{figure}
 \centering\setlength{\unitlength}{0.12cm}
\begin{picture}(70,80)
 \put(40,0){\makebox(0,0)[t]{$b$(fm)}}
 \put(-5,35){\rotatebox[origin=c]{90}{\makebox(0,0)[t]{$P_1(b)$}}}
\put(0,0){\includegraphics[width=70\unitlength,keepaspectratio=true]{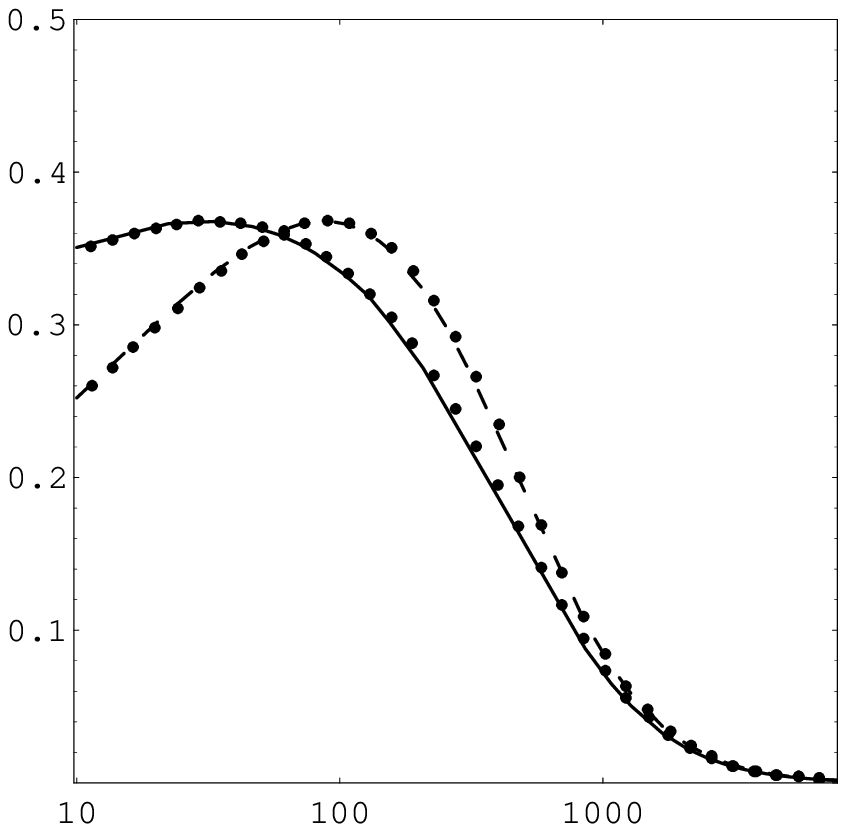}}
\end{picture}
\caption{The probability of one-pair production $P_1(b)$ corresponding to the function
$W(b)$ from Eq. \eqref{eq:Gdefinition}, $\gamma=2\times10^4$, $Z_A=Z_B=79$.
Dashed line: the function $\mathcal{F}$ is taken in the Born approximation,
$\mathcal{F}=\mathcal{F}_0$; solid line: the Coulomb corrections are taken into
account, $\mathcal{F}=\mathcal{F}_0+2\mathcal{F}_A$. Dots show the corresponding
results of numerical calculations from Ref. \cite{HTB99}. }\label{fig:comparison}
\end{figure}

As we already pointed out, Eq. \eqref{Eq:Ftildedef} has logarithmic accuracy which can
be sufficient for very large $\gamma$. In order to go beyond the logarithmic accuracy,
we represent $W(b)$ in the form
\begin{equation}\label{eq:Gdefinition}
W( b)=\zeta\,[L-G(mb)]\,\mathcal{F}(m b)\,,
\end{equation}
where $G(mb)$ is some function of $mb$ and, generally speaking, of the parameters
$Z_A\alpha$ and $Z_B\alpha$. The asymptotics of $G(mb)$ at $1\ll mb\ll \sqrt{\gamma}$
is known, see Eqs. \eqref{WB1} and \eqref{eq:WC1}. However, the calculation of the
function $G(mb)$ at $mb\lesssim 1$ is rather complicated problem. Instead, we use the
results of numerical calculations, performed for a few values of $\gamma$ in Refs.
\cite{HTB95b,HTB99} in the Born approximation. We have found that the form
\begin{equation}\label{eq:Gfit}
G(mb)=\frac32\ln (mb+1)+1.9\,,
\end{equation}
provides good agreement of Eq. \eqref{eq:Gdefinition} with the numerical results of Refs.
\cite{HTB95b,AHTB97,HTB99} in the wide region of $mb$, see Fig. \ref{fig:comparison}.
The form \eqref{eq:Gfit} of $G(mb)$ is obtained by fitting the Born results and thus is
independent of $Z_{A,B}$. It provides the correct asymptotics of $W_0(b)$, Eq.
\eqref{WB1}. It turns out that the formula \eqref{eq:Gdefinition}  with $G(mb)$ from Eq.
\eqref{eq:Gfit} has also high accuracy for $Z_A\alpha,\,Z_B\alpha\lesssim 1$  in the
region $mb\lesssim 1$ where the Coulomb corrections are large. We have checked this
fact by comparing our results with  those  of Ref. \cite{HTB99} obtained numerically for
$Z_A=Z_B=79$, see Fig. \ref{fig:comparison}. Note that the tabulation of $W(b)$ and
$P_N(b)$, performed in Refs.\cite{HTB95b,AHTB97,HTB99,Baltz05,Baltz06} for a few
values of $\gamma$, required the evaluation of nine-fold integral and, therefore, was
very laborious. The calculation of the function $\mathcal{F}$ from Eq. \eqref{eq:Wres2}
is essentially simpler. Besides, since this function is independent of $\gamma$, one can
easily obtain predictions for $W(b)$ at any $\gamma\gg 1$ using Eqs.
\eqref{eq:Gdefinition} and \eqref{eq:Gfit}.

\begin{table}
\begin{tabular}{||r|l|l|l|l||r|l|l|l|l||}
\hline\hline%
$x      $&Born    &Au      &Pb     &   U    &$   x  $&  Born                &             Au       &         Pb          &U                      \\
\hline%------------------------------------------------------------------------------------------------------------------------------------
$ 0.01   $&$ 3.42 $&$ 2.76 $&$2.71 $&$ 2.56 $&$ 1.26  $&$ 0.391              $&$ 0.347              $&$0.343              $&$0.332                $\\
$ 0.0126 $&$ 3.26 $&$ 2.65 $&$2.59 $&$ 2.45 $&$ 1.58  $&$ 0.304              $&$ 0.273              $&$0.27               $&$0.262                $\\
$ 0.0158 $&$ 3.11 $&$ 2.52 $&$2.47 $&$ 2.34 $&$ 2.    $&$ 0.231              $&$ 0.209              $&$0.207              $&$0.202                $\\
$ 0.02   $&$ 2.96 $&$ 2.4  $&$2.35 $&$ 2.22 $&$ 2.51  $&$ 0.171              $&$ 0.156              $&$0.155              $&$0.152                $\\
$ 0.0251 $&$ 2.8  $&$ 2.28 $&$2.24 $&$ 2.11 $&$ 3.16  $&$ 0.124              $&$ 0.114              $&$0.114              $&$0.111                $\\
$ 0.0316 $&$ 2.65 $&$ 2.16 $&$2.12 $&$ 2.   $&$ 3.98  $&$ 8.78\times 10^{-2} $&$ 8.2 \times 10^{-2} $&$8.15 \times 10^{-2}$&$8.01 \times 10^{-2}  $\\
$ 0.0398 $&$ 2.5  $&$ 2.04 $&$2.0  $&$ 1.89 $&$ 5.01  $&$ 6.14\times 10^{-2} $&$ 5.78 \times 10^{-2}$&$5.75 \times 10^{-2}$&$5.66 \times 10^{-2}  $\\
$ 0.0501 $&$ 2.34 $&$ 1.92 $&$1.88 $&$ 1.78 $&$ 6.31  $&$ 4.25\times 10^{-2} $&$ 4.02 \times 10^{-2}$&$4.0  \times 10^{-2}$&$3.95 \times 10^{-2}  $\\
$ 0.0631 $&$ 2.19 $&$ 1.8  $&$1.76 $&$ 1.67 $&$ 7.94  $&$ 2.91\times 10^{-2} $&$ 2.77 \times 10^{-2}$&$2.76 \times 10^{-2}$&$2.73 \times 10^{-2}  $\\
$ 0.0794 $&$ 2.04 $&$ 1.68 $&$1.64 $&$ 1.56 $&$ 10    $&$ 1.99\times 10^{-2} $&$ 1.9 \times 10^{-2} $&$1.89 \times 10^{-2}$&$1.87 \times 10^{-2}  $\\
$ 0.1    $&$ 1.88 $&$ 1.55 $&$1.52 $&$ 1.45 $&$ 12.6  $&$ 1.35\times 10^{-2} $&$ 1.29 \times 10^{-2}$&$1.29 \times 10^{-2}$&$1.28 \times 10^{-2}  $\\
$ 0.126  $&$ 1.73 $&$ 1.43 $&$1.41 $&$ 1.34 $&$ 15.8  $&$ 9.07\times 10^{-3} $&$ 8.75 \times 10^{-3}$&$8.72 \times 10^{-3}$&$8.64 \times 10^{-3}  $\\
$ 0.158  $&$ 1.58 $&$ 1.31 $&$1.29 $&$ 1.23 $&$ 20    $&$ 6.09\times 10^{-3} $&$ 5.89 \times 10^{-3}$&$5.87 \times 10^{-3}$&$5.83 \times 10^{-3}  $\\
$ 0.2    $&$ 1.43 $&$ 1.19 $&$1.17 $&$ 1.12 $&$ 25.1  $&$ 4.07\times 10^{-3} $&$ 3.95 \times 10^{-3}$&$3.94 \times 10^{-3}$&$3.91 \times 10^{-3}  $\\
$ 0.251  $&$ 1.28 $&$ 1.07 $&$1.06 $&$ 1.01 $&$ 31.6  $&$ 2.71\times 10^{-3} $&$ 2.64 \times 10^{-3}$&$2.63 \times 10^{-3}$&$2.61 \times 10^{-3}  $\\
$ 0.316  $&$ 1.14 $&$ 0.961$&$0.941$&$ 0.898$&$ 39.8  $&$ 1.8 \times 10^{-3} $&$ 1.75 \times 10^{-3}$&$1.75 \times 10^{-3}$&$1.74 \times 10^{-3}  $\\
$ 0.398  $&$ 0.993$&$ 0.842$&$0.829$&$ 0.793$&$ 50.1  $&$ 1.19\times 10^{-3} $&$ 1.16 \times 10^{-3}$&$1.16 \times 10^{-3}$&$1.15 \times 10^{-3}  $\\
$ 0.501  $&$ 0.856$&$ 0.731$&$0.72 $&$ 0.69 $&$ 63.1  $&$ 7.9 \times 10^{-4} $&$ 7.71 \times 10^{-4}$&$7.69 \times 10^{-4}$&$7.65 \times 10^{-4}  $\\
$ 0.631  $&$ 0.725$&$ 0.625$&$0.616$&$ 0.591$&$ 79.4  $&$ 5.21\times 10^{-4} $&$ 5.09 \times 10^{-4}$&$5.08 \times 10^{-4}$&$5.05 \times 10^{-4}  $\\
$ 0.794  $&$ 0.603$&$ 0.524$&$0.517$&$ 0.498$&$ 100   $&$ 3.43\times 10^{-4} $&$ 3.35 \times 10^{-4}$&$3.34 \times 10^{-4}$&$3.33 \times 10^{-4}  $\\
$ 1.     $&$ 0.491$&$ 0.431$&$0.426$&$ 0.411$&&&&&\\
 \hline
\end{tabular}
\caption{The function $\mathcal{F}(x)$, Eq. \eqref{eq:Wres2}, calculated in the Born
approximation ($Z_A\alpha\to 0$) and exactly in the parameter $Z_A\alpha$ for Au, Pb,
and U.} \label{tab:Ffunctions}
\end{table}

\section{Conclusion}
In the present paper, we have found in the leading logarithmic approximation the simple
representation for the  function $W(b)$ for $mb\lesssim 1$, $Z_B\alpha\ll 1$, and
arbitrary $Z_A\alpha$. Using the results of numerical calculation of $W(b)$ performed for
a few values of $\gamma$ and $Z_{A,B}$, we have obtained the approximate formula for
$W(b)$ valid in a wide region of parameters: $mb\lesssim \sqrt{\gamma}$,
$Z_A\alpha\lesssim 1$, $Z_B\alpha\lesssim 1$, $\gamma\gg 1$. We estimate the
accuracy of this formula to be a few percent. The results obtained clearly demonstrate
the dependence of $W(b)$, as well as $P_n(b)$, on the relativistic factor $\gamma$ and
the parameters $Z_{A,B}\alpha$.

This work was supported in part by RFBR grant No. 05-02-16079 and by the grant for
young scientists of SB RAS (R.N.L.).

\appendix

\section{Calculation of the integrals}
%%%%%%%%%%%%%%%%%%%%%%%%%%%%%%%%
In this Appendix, we present some details of derivation of Eq. \eqref{eq:Wres1} from Eq.
\eqref{dW2}. The main contribution to the integrals comes from the region of small angles
between vectors $\bm{r}_1$,  $-\bm{r}_2$, and $z$-axis. Using this fact, we take the
integrals over the angles of $\bm{r}_1$ and $\bm{r}_2$, make the substitution $r_{1,2}\to
E\,r_{1,2}$, and change the variables $\varepsilon_p=E x$, $\varepsilon_q=E
\bar{x}=E(1-x)$. Taking the integral over $E$ in the logarithmic approximation at
$\gamma\gg 1$ and $mb\lesssim 1$, we obtain

\begin{align}
dW(b)  &  =\frac{\left(  Z_B\alpha\right)  ^{2}}{\left(
2\pi\right)  ^{6}}\ln\gamma\int\frac{d\bm{k}_{1\bot}}{k%
_{1\bot}^{2}}\frac{d\bm{k}_{2\bot}}{k_{2\bot}^{2}}\int dx\,x\,\bar
{x}\frac{dr_{1}}{r_{1}}\frac{dr_{2}}{r_{2}}\int d\bm{Q\,}%
d\bm{q} \left(\frac{\left\vert\bm{q}+\bm{Q}\right\vert}{ \left\vert\bm{q}
-\bm{Q}\right\vert}\right) ^{2iZ_A\alpha}\nonumber\\
&  \times\exp\left[  -\frac{i}{2}m^{2}(r_1+r_2)-i\,\bm{\Delta}\cdot
\bm{\beta}-\frac{i}{2}x\bar{x}(r_{1}k_{1\bot}^{2}%
+r_{2}k_{2\bot}^{2})+\frac{i(r_1+r_2)\bm{Q}^{2}}{2r_{1}r_{2}}\right]
\nonumber\\
&  \times\left\langle 2\left(  \bar{x}-x\right)  \left(
\frac{k_{1\bot}^{2}\bm{k}_{2\bot}\cdot\bm{Q}}{r_{2}}
-\frac{k_{2\bot}^{2}\bm{k}_{1\bot}\cdot\bm{Q}}{r_{1}}\right)
-4x\bar
{x}k_{1\bot}^{2}k_{2\bot}^{2}-\frac{4\left(  \bm{k}%
_{1\bot}\cdot\bm{Q}\right)  \left(  \bm{k}_{2\bot}\cdot\bm{Q}%
\right)  }{r_{1}r_{2}}\right. \nonumber\\
&  -\left.  \left(  \bm{k}_{1\bot}\cdot\bm{k}_{2\bot}\right) \left(
\frac{m^{2}(r_1+r_2)^{2}}{2x\bar{x}r_{1}r_{2}}+\frac{r_{1}k_{1\bot}^{2}%
}{2r_{2}}+\frac{r_{2}k_{2\bot}^{2}}{2r_{1}}\right)  \right\rangle
,\nonumber\\
\bm{\Delta}  &  =\bm{k}_{1\bot}%
-\bm{k}_{2\bot},\quad\beta=\bm{q}/2+%
\left(  \bar{x}-x\right) \bm{Q}/2-\bm{b}\,.\label{dWquasi}
\end{align}
The integration over two-dimensional vectors $\bm{k}_{1\bot}$ and $\bm{k}_{2\bot}$
can be easily performed. The result reads
\begin{align}
dW(\boldsymbol{b})  &  =-\frac{\left(  Z_B\alpha\right)  ^{2}}{\left(
2\pi\right)  ^{4}}\ln\gamma\int dx\,\frac{dr_{1}}{r_{1}^{2}}\frac{dr_{2}%
}{r_{2}^{2}}\int d\boldsymbol{Q\,}d\boldsymbol{q}\exp\left[  -\frac{i}{2}m^{2}(r_1+r_2)+\frac{i(r_1+r_2)\boldsymbol{Q}^{2}}%
{2r_{1}r_{2}}\right] \nonumber\\
&  \times\left(\frac{\left\vert\bm{q}+\bm{Q}\right\vert}{ \left\vert\bm{q}
-\bm{Q}\right\vert}\right) ^{2iZ_A\alpha}\left\langle \left[  2\left(  \bar{x}-x\right)
\frac {\boldsymbol{\beta}\cdot\boldsymbol{Q}}{\boldsymbol{\beta}^{2}}+\frac
{1}{2 x\bar{x}}\right]  \left(  2\mathcal{E}_{1}\mathcal{E}_{2}-\mathcal{E}_{1}%
-\mathcal{E}_{2}\right)  -4\mathcal{E}_{1}\mathcal{E}_{2}\right. \nonumber\\
&  +\left.  \left[  \frac{4\left(  \boldsymbol{\beta}\cdot\boldsymbol{Q}%
\right)  ^{2}}{\boldsymbol{\beta}^{4}}x\bar{x}+\frac{m^{2}(r_1+r_2)^{2}}%
{2\boldsymbol{\beta}^{2}}\right]  \left(  \mathcal{E}_{1}-1\right)  \left(
\mathcal{E}_{2}-1\right)  \right\rangle ,\label{dxdr1dr2dQdq}\\
\mathcal{E}_{i}  &  =\exp[i\boldsymbol{\beta}^{2}/2x\bar{x}r_{i}]\nonumber
\end{align}
Taking the integrals over $r_{1,2}$ and passing from the variable $\bm{q}$ to
$\bm{\beta}=\bm{q}/2+ \left(  \bar{x}-x\right) \bm{Q}/2-\bm{b}$, we obtain Eq.
\eqref{eq:Wres1}.

\end{document}